\documentclass[aps,prb,twocolumn,showpacs,groupedaddress,amsmath,a4paper]{revtex4}

\usepackage{graphicx,bm,enumerate,hyphenat}

\begin{document}

\title{SU(2) approach to the pseudogap phase of high-temperature superconductors: electronic spectral functions}

\author{Samuel Bieri}
\email[]{samuel.bieri@a3.epfl.ch}
\author{Dmitri A.\ Ivanov}
\affiliation{
Institute of Theoretical Physics,
Ecole Polytechnique F\'{e}d\'{e}rale de Lausanne (EPFL),
CH-1015 Lausanne,
Switzerland
}


\begin{abstract}

We use an SU(2) mean-field theory approach with input from variational wavefunctions of the $t\text{-}J$ model to study the electronic spectra in the pseudogap phase of cuprates. In our model, the intermediate-temperature state of underdoped cuprates is realized by classical fluctuations of the order parameter between the $d$-wave superconductor and the staggered-flux state. The spectral functions of the pure and the averaged states are computed and analyzed. Our model predicts a photoemission spectrum with an asymmetric gap structure interpolating between the superconducting gap centered at the Fermi energy and the asymmetric staggered-flux gap. This asymmetry of the gap changes sign at the points where the Fermi surface crosses the diagonal $(\pi,0)\text{-}(0,\pi)$.

\end{abstract}
\maketitle

The most unusual and debated feature of high-temperature superconductivity~(HTSC) is the pseudogap~(PG) phase, the high-temperature phase in the underdoped region of the phase diagram between the destruction of superconductivity at $T_c$ and the pseudogap temperature $T^*$.\cite{timusk99,norman05} While the zero-temperature phase diagram of HTSC is relatively well understood, there is currently much experimental and theoretical interest in the intermediate-temperature PG phase. In this phase, several surprising experimental features appear: e.g.\ angle-resolved photoemission spectroscopy~(ARPES) shows a state which is partially gapped on the experimental Fermi surface.\cite{marshall96,loessner96,ding96,norman97,kanigel07,campuzano04,damascelli03}

Theoretically, the low-temperature physics of HTSC is well described by variational wavefunctions of the $t\text{-}J$ model.\cite{anderson04,leeReview08,grosReview07,ogataReview08,hasegawa89,giamarchiLhuillier93,gros88} The antiferromagnetic parent state at half filling is destroyed as doping is increased. Due to the gain of spin-exchange energy in the Gutzwiller-projected state, a $d$-wave mean-field order is favored away from half filling. The characteristic dome for the off-diagonal order can be reproduced variationally.\cite{paramekanti0103,bieri07} Low-lying Gutzwiller-projected quasiparticle excitations reproduce well many experimental features.\cite{paramekanti0103, sensarma05, yunoki05, yunoki05prl, yunoki06, yanChouLee06, chou06, bieri07,RanWen06} The main disadvantage of the variational approach is that it is a zero-temperature theory and cannot easily be extended to finite temperature or to high-energy excitations.\cite{grosReview07}

Many years ago, it was noticed that there is a redundant description of Gutzwiller-projected fermionic wavefunctions exactly at half filling, parametrized by local SU(2) rotations.\cite{affleck88,dagotto88} Away from half filling, this redundancy is lifted. Later, Wen and Lee {\it et al.} proposed a slave-boson field theory, where the redundancy is promoted to a dynamical SU(2) gauge theory away from half filling.\cite{wenlee96,lee98,rantnerWen00}
The advantage of the SU(2) slave-boson approach is that it incorporates strong correlations when gauge fluctuations around the mean-field saddle points are included. Integrating over all gauge-field configurations in this approach enforces the Gutzwiller constraint $n_i<2$. The slave-boson mean-field theory is then not restricted to low temperatures.

The SU(2) approach to the $t\text{-}J$ model predicts that a state with staggered magnetic fluxes through the plaquettes of the square lattice should be close in energy to the $d$-wave superconductor at low doping.\cite{wenlee96,lee98,leeWen01,leeProceedings} In fact, a staggered SU(2) rotation on nearest-neighbor sites transforms the $d$-wave superconductor~(SC) into the staggered-flux~(SF) state.
These two states are identical at half filling. At small doping, one expects the local symmetry to be weakly broken, and the SU(2) rotation provides a route to construct a low-lying nonsuperconducting variational state for the weakly doped $t\text{-}J$ model. This led to the proposal by Wen and Lee that the pure SF state should be realized in the vortex cores of HTSC.\cite{leeWen01} Indeed, it was confirmed numerically that the Gutzwiller-projected SF state is a very competitive ground state of the $t\text{-}J$ model.\cite{ivanovLee03} Further support for the SU(2) approach came from the discovery of SF correlations in the Gutzwiller-projected $d$-wave superconductor.\cite{ivanovLeeWen00}

In this paper, we restrict ourselves to the so-called ``staggered $\theta$-mode'' which interpolates between the SC and the SF states.\cite{leeNagaosa03} As the temperature is increased through $T_c$ in the underdoped compounds, vortices proliferate and eventually destroy the phase coherence. In order to form energetically inexpensive vortices in the superconductor, the order parameter rotates to the SF state inside the cores.\cite{kishineLeeWen01} However, in contrast to vortex cores, we do not expect a pure SF state to be realized in the bulk. The PG state should be viewed as a thermal average over different intermediate states between the SF and the SC state, parametrized by appropriate SU(2) rotations.

In the superconducting phase at low temperature, it is sufficient to include Gaussian fluctuations away from the superconducting state. In this framework, Honerkamp and Lee found that coupling to the Gaussian $\theta$-mode strongly depletes the antinodal quasiparticles.\cite{honerkampLee03} This is in contrast to zero temperature, where Gutzwiller-projected excitations show rather weak reduction of spectral weight in the antinodal region as shown by the authors.\cite{bieri07} At temperatures between $T_c$ and $T^*$, strong fluctuations toward the SF state are expected to affect the electronic spectral functions even more. In the present work, we are interested in the electron spectral intensities in the pseudogap region, i.e.\ in the presence of large fluctuations of the order parameter between the SC and the SF states.

Our model bears some similarity to the $\sigma$-model approach for the SU(2) gauge theory of the $t\text{-}J$ model, introduced by Lee {\it et al.}\cite{lee98,leeReview08} In contrast to these authors, we do not use a self-consistent mean-field treatment, but we consider an effective model with input from Gutzwiller-projected variational wavefunctions of the $t\text{-}J$ model.


A complementary study was conducted by Honerkamp and Lee who considered SU(2)~fluctuations in an inhomogeneous vortex liquid.\cite{honerkampLee04} These authors computed the density of states and helicity modulus, and found that a dilute liquid of SF vortices would account for the large Nernst signal observed in the pseudogap phase.\cite{nernstSignal} In the present paper, we are particularly interested in the implications of the fluctuating-staggered-flux scenario for the ARPES spectra.

Finally, let us note that our model concerns the low-energy spectra of cuprate superconductors, $|\omega|\lesssim 200\,{\rm meV}$. The interesting high-energy anomalies ($|\omega|\simeq 0.4 \text{ - }1\,{\rm eV}$) which were discovered in recent experimental\cite{lanzara08,zhang08} and theoretical\cite{tanWang08} works are not in the scope of the current discussion.

This paper is organized in the following way. In Sec.~\ref{sec:model}, we introduce the model and describe the observable (spectral function) that we want to study. In Sec.~\ref{sec:pure}, we give a detailed account on the spectra of the pure (unaveraged) states. Finally, in Sec.~\ref{sec:avg} we present our results on the spectral functions averaged over the order-parameter space and in Sec.~\ref{sec:exp} we discuss the experimental implications.

\section{Model}\label{sec:model}

The local SU(2) rotation for the $t\text{-}J$ model\cite{leeReview08} is conveniently written using spinon doublets in the usual notation,
${\bm\psi}^\dagger=(c_\uparrow,c_\downarrow^\dagger)$. In terms of these doublets, the SF state is defined by the mean-field Hamiltonian $H_{SF} = \sum_{\langle i,j\rangle} {\bm\psi}_i^\dagger U^{SF}_{ij} {\bm\psi}_j -\mu \sum_i {\bm\psi}_i^\dagger\sigma_3 {\bm\psi}_i$ where $U^{SF}_{ij} = -\chi\sigma_3 - i \Delta (-)^{i_x+j_y}$, $\sigma_\alpha$ are the Pauli matrices, and the sum $\langle i,j \rangle$ is taken over pairs of nearest-neighbor sites. In this paper, we restrict ourselves to the following SU(2) rotation: $U_{ij} \rightarrow g_i^\dagger U_{ij} g_j$, with $g_j=\exp[i (-)^{j_x+j_y} \frac{\theta}{2}\, \sigma_1]$. Note that for $\theta=\frac{\pi}{2}$, the SF Hamiltonian is rotated to a $d$-wave superconductor, $U^{SF}_{ij}\rightarrow U^{SC}_{ij} = -\chi\sigma_3 + \Delta (-)^{i_x+j_x} \sigma_1$. $U^{SC}$ represents the $d$-wave mean field with pairing $\Delta$. $U^{SF}$ is the nonsuperconducting mean field with staggered U(1) fluxes equal to $\pm 4 \arctan(\Delta/\chi)$ through each plaquette of the square lattice. The intermediate states for general $\theta$ contain both $d$-wave pairing and staggered fluxes.

We now consider the mean-field Hamiltonian at the intermediate values of $\theta$ between $0$ and $\pi$:
\begin{equation}\begin{split}
H_{MF}(\theta) = &\sum_{\langle i,j\rangle} {\bm\psi}_i^\dagger g_i^\dagger(\theta) U_{ij}^{SF} g_j(\theta) {\bm\psi}_j^\dagger \\
& - \chi' \sum_{\langle i,j\rangle'} {\bm\psi}_i^\dagger \sigma_3 {\bm\psi}_j - \mu \sum_i {\bm\psi}_i^\dagger \sigma_3 {\bm\psi}_i\, .
\end{split}\label{eq:Hmf}\end{equation}
As usual, the chemical potential $\mu$ is added to enforce the desired average particle number. We have also added a phenomenological next-nearest-neighbor hopping $\chi'$.
%
%
%
Note that the parameters $\chi$, $\chi'$, and $\Delta$ of Hamiltonian (\ref{eq:Hmf}) are the effective parameters describing the variational ground state and quasiparticle spectrum of the $t\text{-}J$ model, for the physically relevant value $t\simeq 3J$: The hopping $\chi$ only weakly depends on doping (at small doping) and is approximately given by $\chi\simeq t/3 \simeq 100\, {\rm meV}$.\cite{yunoki05prl,edeggerAnderson06,paramekanti0103} At $10\%$ doping,  $\Delta$ decreases slightly from $\Delta_{\theta=\pi/2} \simeq 0.25\chi$ in the SC state to $\Delta_{\theta=0} \simeq 0.2\chi$ in the SF state.\cite{ivanovLee03,note_params}

The value of the next-nearest-neighbor hopping is taken to be $\chi'=-0.3\chi$, to mimick the experimental Fermi surface observed in cuprates. Earlier studies of Gutzwiller-projected wavefunctions suggest that such an effective next-nearest-neighbor hopping may appear in the underdoped region as a consequence of strong correlations, even in the absence of the term in the physical Hamiltonian.\cite{bieri07,himedaOgata00} Note that we keep this term unrotated in Eq.~\eqref{eq:Hmf}.

In our model, physical quantities at finite temperature are given by an appropriate functional integral over the mean-field parameters $U_{ij}$, weighted by a free energy which is almost flat in the directions parametrized by $g_j$. As indicated earlier, we restrict our study to staggered SU(2) rotations parametrized by the angle $\theta$. At the same time, we neglect the amplitude fluctuations of $\Delta$, since the energy scale associated with these fluctuations is high: of the order of $T^*$, in our approach. On the other hand, the energy scale $\varepsilon_c = E_{SF}-E_{SC}$, responsible for the $\theta$-fluctuations, is much lower: at 10\% doping it is estimated as $\varepsilon_c \simeq 0.02 J \simeq 30 K$ (per lattice site) from variational Monte Carlo calculations.\cite{ivanovLee03}

The free energy describing classical fluctuations of $\theta$ contains a $\theta$-dependent ``condensation energy'' (of the order $\varepsilon_c$) and a gradient term\cite{honerkampLee04} $\rho ({\bm \nabla} \theta)^2$. We assume a situation where the resulting correlation length $\xi = \sqrt{\rho/\varepsilon_c}$ is much larger than one lattice spacing.\cite{note_rho,cohlength} In this case, the characteristic temperature, below which the condensation energy selects the superconducting state over the staggered-flux one, is $T_c \sim \rho/\ln\xi$ (the same scale determines the temperature of the Kosterlitz-Thouless-type transition).\cite{okwamoto84} For temperatures above that scale, but below $\rho$,
\begin{equation}
  \rho/\ln\xi < T < \rho\,,
\end{equation}
the order parameter slowly varies in space and takes all possible values related by the SU(2) rotation. Therefore, in this temperature range, we can approximate
%
the classical fluctuations by an equal-weight statistical average over the uniform states with all possible values of $\theta$. The corresponding integration measure for $\theta$ is $\int_0^1 d(\cos\theta)$, inherited from the invariant measure on SU(2). 

We calculate the spectral function $A^{\theta}_{\bm k}(\omega)=-\frac{1}{\pi} \text{Im}G_{\bm k}(\omega+i\Gamma)$ where $G$ is the single-particle Green's function\cite{note_gf} of $H_{MF}$, Eq.~\eqref{eq:Hmf}. Note that $\omega$ is measured with respect to the Fermi energy throughout this paper. As explained above, the spectra in the pseudogap phase are modeled by the averages of this spectral function over the order-parameter space, $A_{\bm k}(\omega)=\int_0^1 d(\cos\theta) A^{\theta}_{\bm k}(\omega)$. The spectral functions of the pure states [$A_{\bm k}^{\theta}(\omega)$] are sums of delta functions. After averaging over $\theta$, the spectral functions acquire an intrinsic width. In addition, we introduce a lifetime broadening $\Gamma$ to make the figures more readable.

\begin{figure}[h!t]
\begin{minipage}[tl]{4.25cm} \includegraphics[width=1.65in]{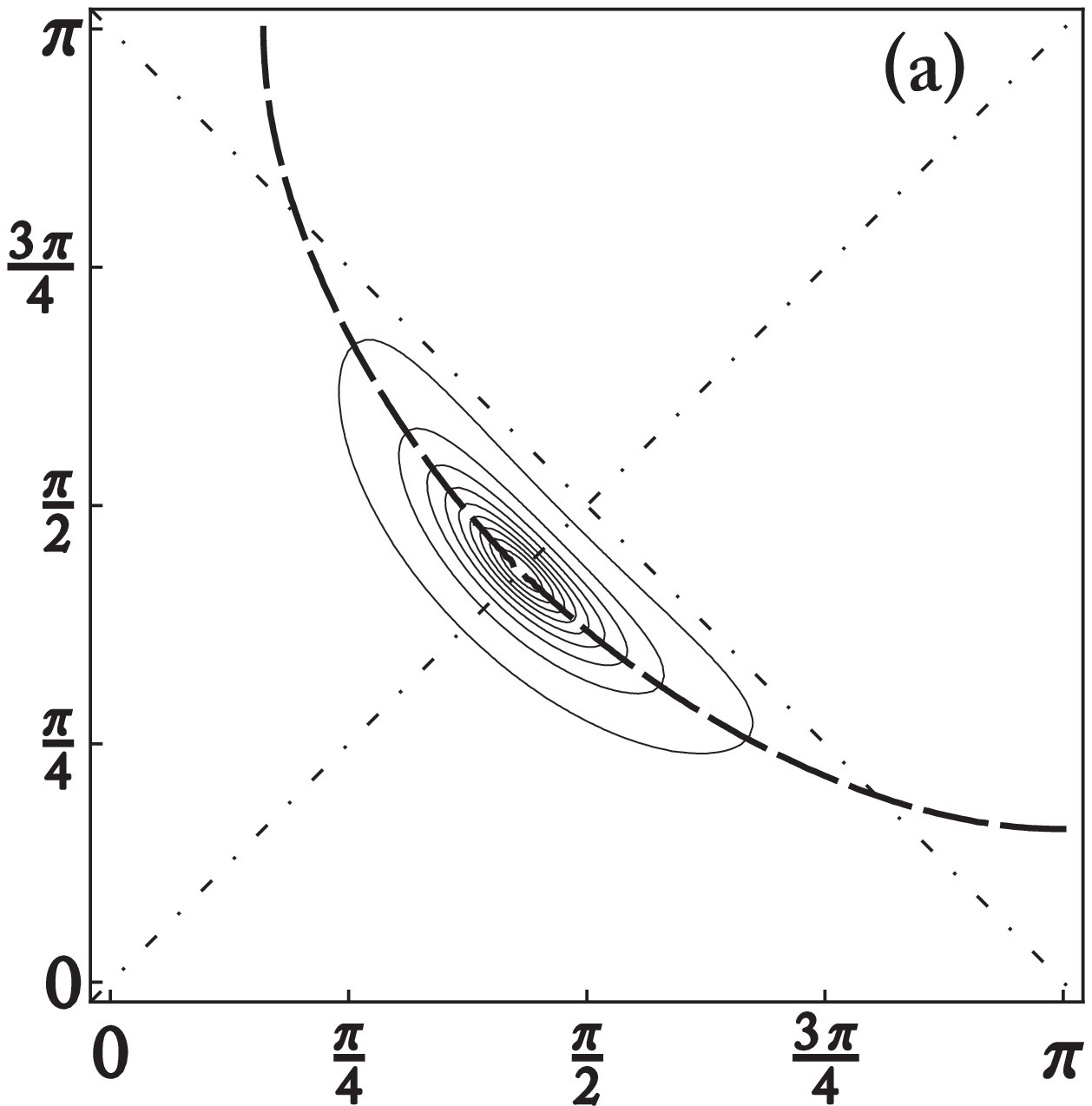} \end{minipage}
\begin{minipage}[tr]{4.25cm} \includegraphics[width=1.65in]{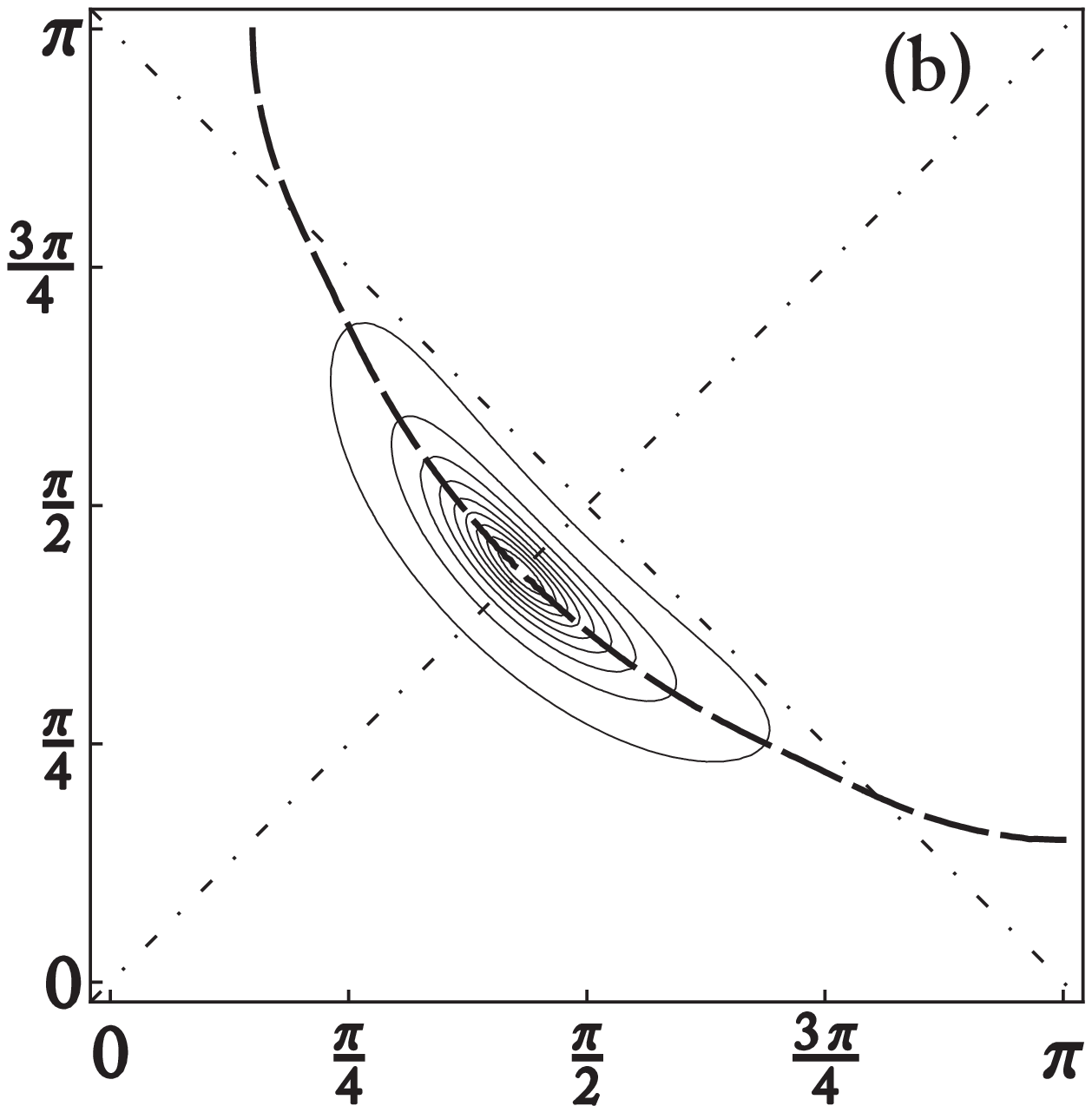} \end{minipage}
\begin{minipage}[bl]{4.25cm} \includegraphics[width=1.65in]{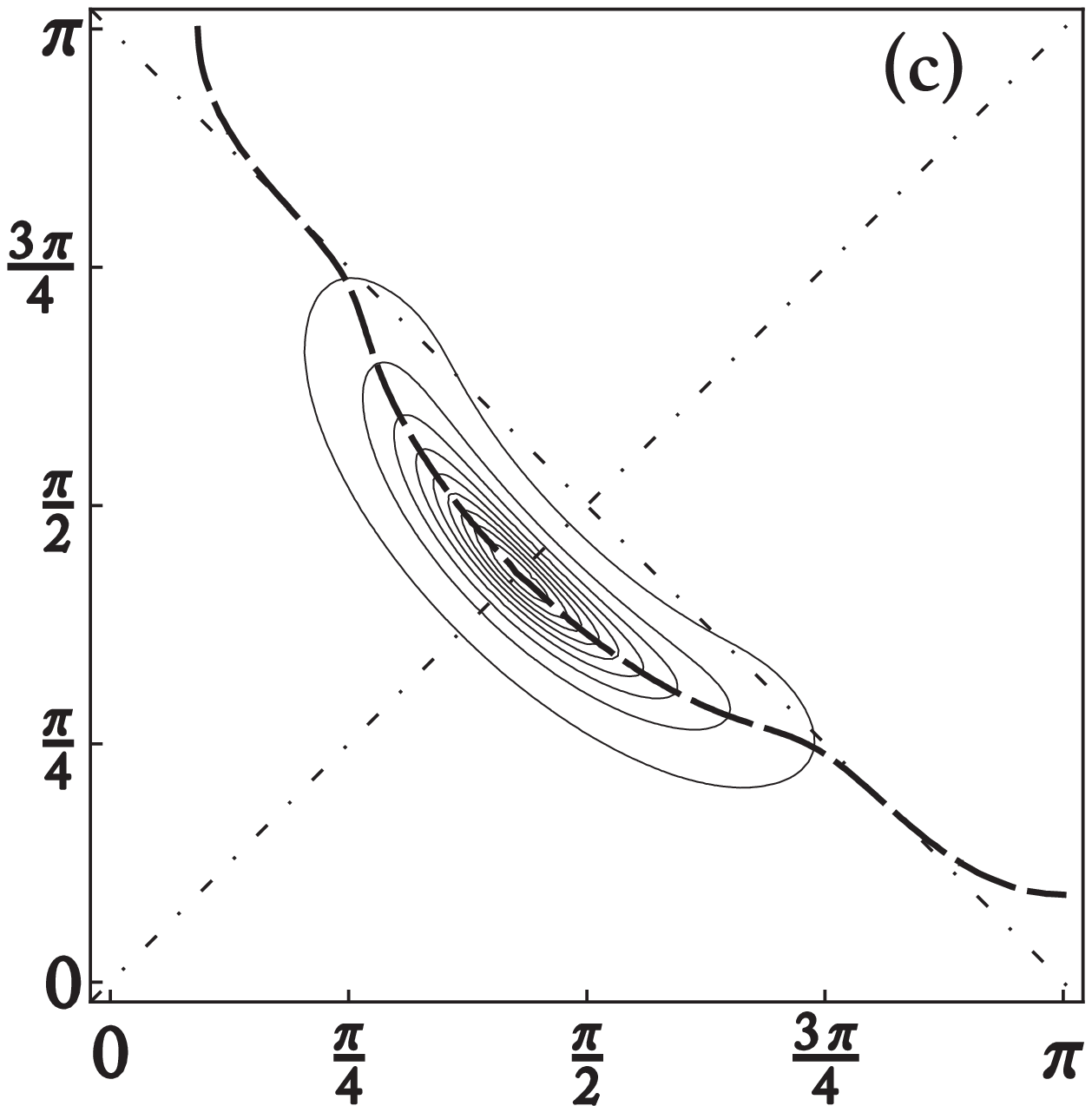} \end{minipage}
\begin{minipage}[br]{4.25cm} \includegraphics[width=1.65in]{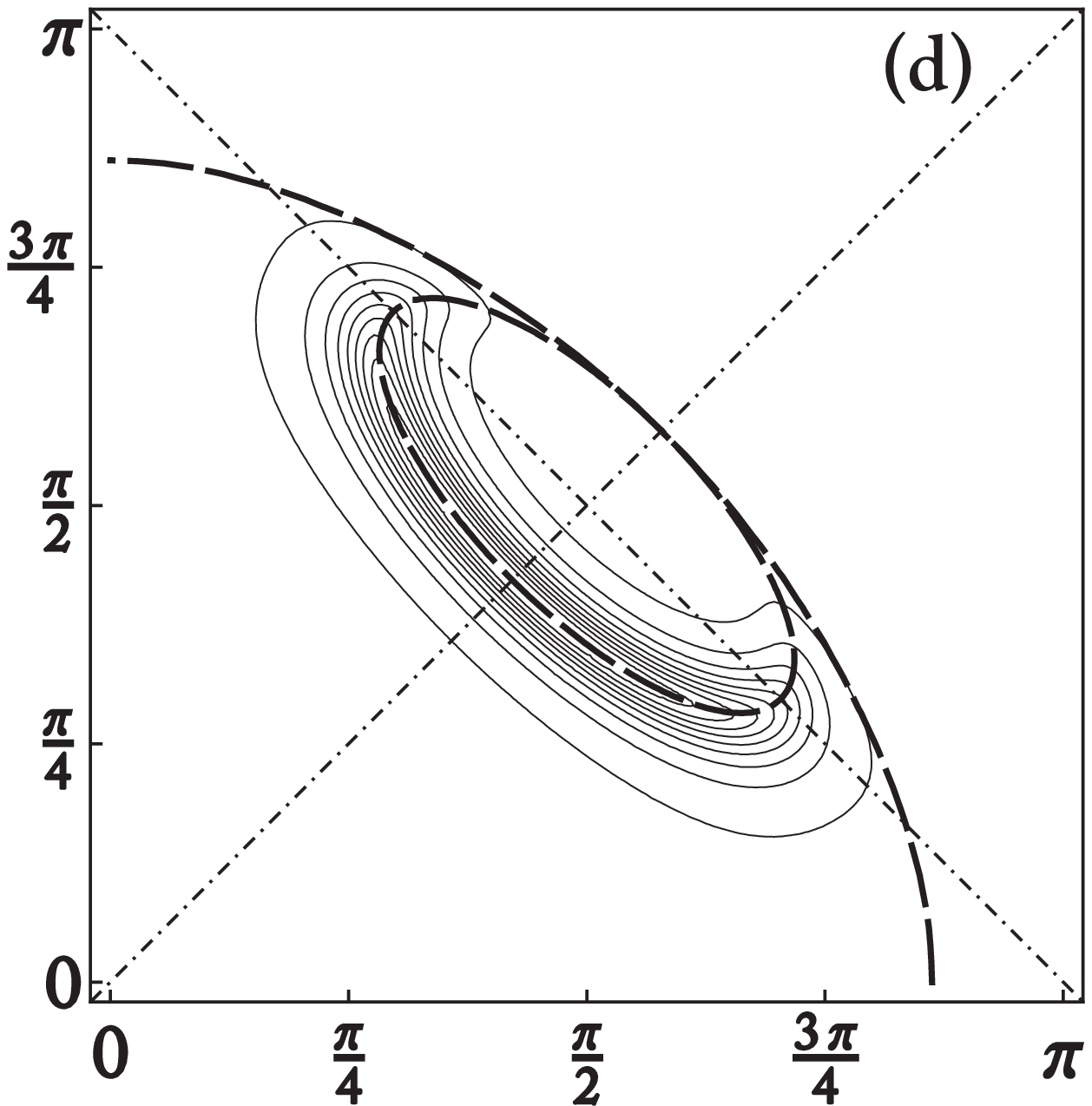} \end{minipage}
\caption{Contour plot of the spectral function at the Fermi energy of the pure states, $A_{\bm k}^\theta(0)$. Doping is $10\%$ and we use a lifetime broadening $\Gamma=0.2\chi$. The dashed line represents the location of the Fermi surface (where the Green's function changes sign\cite{note_FS}). From (a) to (d) we have $\cos\theta = 0, 1/3, 2/3, 1$. Plot (a) shows the $d$-wave superconductor, (d) displays the pure staggered-flux state.}
\label{fig:pureLS}
\end{figure}

\section{Pure states}\label{sec:pure}

In order to understand the averaged spectral function, we first outline how the intermediate states evolve as $\theta$ is decreased from $\theta=\frac{\pi}{2}$ (SC state) to $\theta=0$ (SF state). In Fig.~\ref{fig:pureLS}, we plot the Fermi surface (more precisely the Luttinger surface\cite{note_FS}) and the spectral function at the Fermi energy, $A_{\bm k}^\theta(0)$.
As the parameter $\theta$ is decreased from $\frac{\pi}{2}$, the superconducting Fermi surface gradually deforms to the well-known pocket around $(\frac{\pi}{2},\frac{\pi}{2})$ of the SF state. However, the points on the superconducting Fermi surface where it crosses the diagonal $(\pi,0)$-$(0,\pi)$ do not move as $\theta$ is changed. We will call them {\it SU(2)~points}, because at these points, the full SU(2) symmetry is intact even away from half filling. We will comment more on this later. As we decrease $\theta$, the SC gap, symmetric with respect to the Fermi level, decreases and closes at $\theta=0$ [$\Delta_{SC}=2(\cos k_x-\cos k_y)\,\Delta\,\sin\theta$]. At the same time, an SF gap opens on the diagonal $(\pi,0)$-$(0,\pi)$ at the energy $\omega\simeq -\tilde \mu$ [we define $\tilde\mu=\mu-2\chi'\cos k_x\cos k_y$]. The SF gap value is $\Delta_{SF}\simeq 2(\cos k_x-\cos k_y)\,\Delta\, \cos\theta$. The spectral weight is transferred among the four bands and all of them gain intensity in the intermediate states. However, in most parts of the zone, there is only a single strong band.

\begin{figure}[h!t]
\begin{minipage}[t] {.5\textwidth} \includegraphics[width=2.9in]{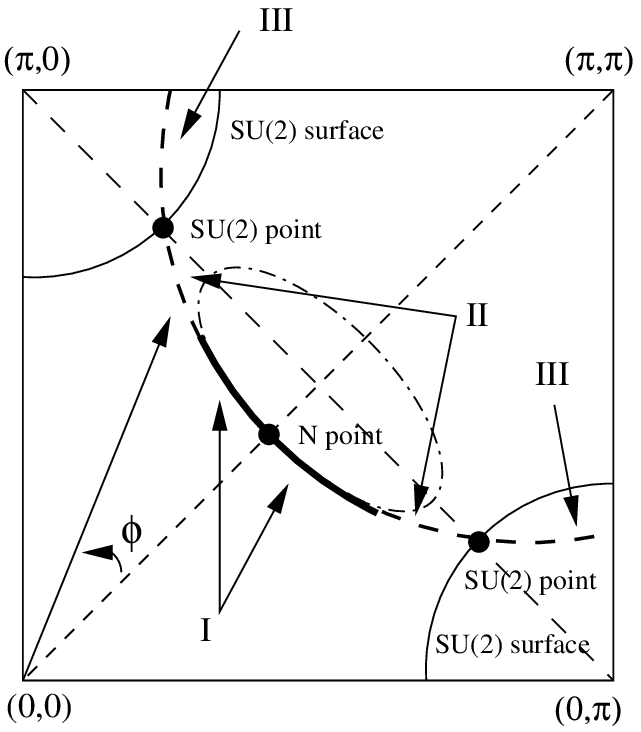} \end{minipage}
\begin{minipage}[b] {.5\textwidth} \includegraphics[width=2.9in]{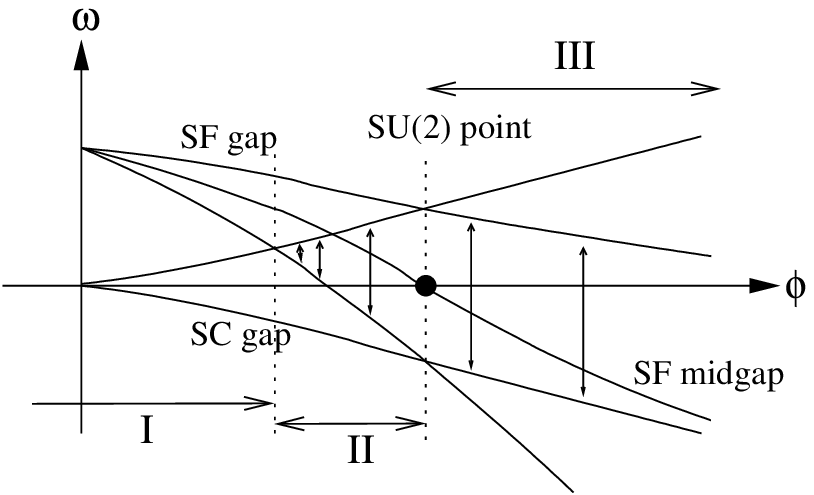} \end{minipage}
\caption{Scheme of the different regions of the Fermi surface. The SC and SF gaps open near the node (N~point; $\phi=0$) where they are small and do not overlap. The Fermi surface appears as a gapless arc in region~I. In region~II, the two gaps start to overlap and form an effective gap which is shifted upward in energy (vertical arrows). The effective gap comes down in energy as we move toward the antinode in region~II. Exactly at the SU(2)~points on the diagonal $(0,\pi)$-$(\pi,0)$, the effective gap is symmetric. Beyond the SU(2)~points (region~III), the midgap is shifted below the Fermi energy.}
\label{fig:FSscheme}
\end{figure}

The SU(2)~points mentioned before belong in fact to SU(2) surfaces (see illustration in Fig.~\ref{fig:FSscheme}) where $\tilde\mu = \mu-2\chi'\cos k_x\cos k_y = 0$. On these surfaces, the full SU(2) symmetry is intact even away from half filling, in the sense that the mean-field spectra are degenerate and independent of $\theta$ (if we neglect the weak dependency of $\Delta$ on $\theta$; see Sec.~\ref{sec:model} and Remark~\onlinecite{note_params}).

\begin{figure}[htbp]
\begin{minipage}[tl]{4.25cm} \includegraphics[width=1.4in]{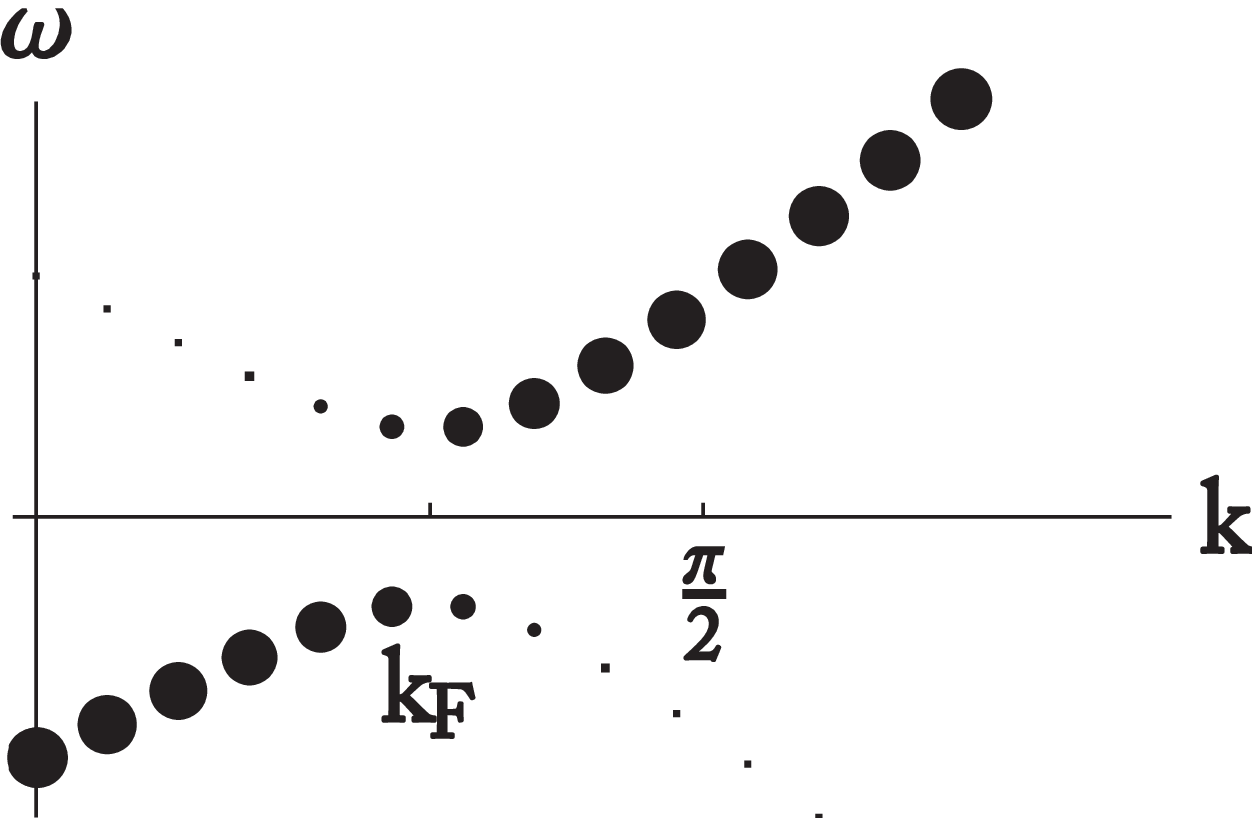} \end{minipage}
\begin{minipage}[tr]{4.25cm} \includegraphics[width=1.4in]{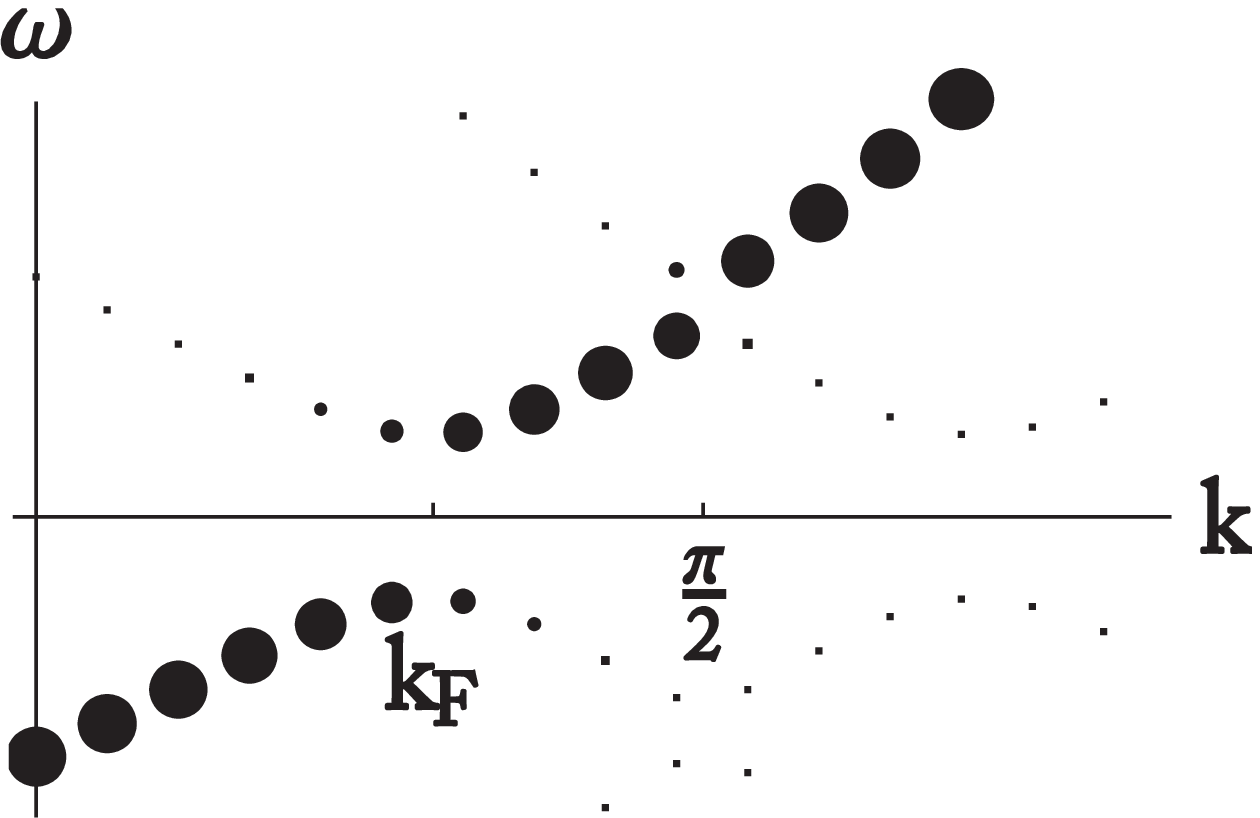} \end{minipage}
\begin{minipage}[bl]{4.25cm} \includegraphics[width=1.4in]{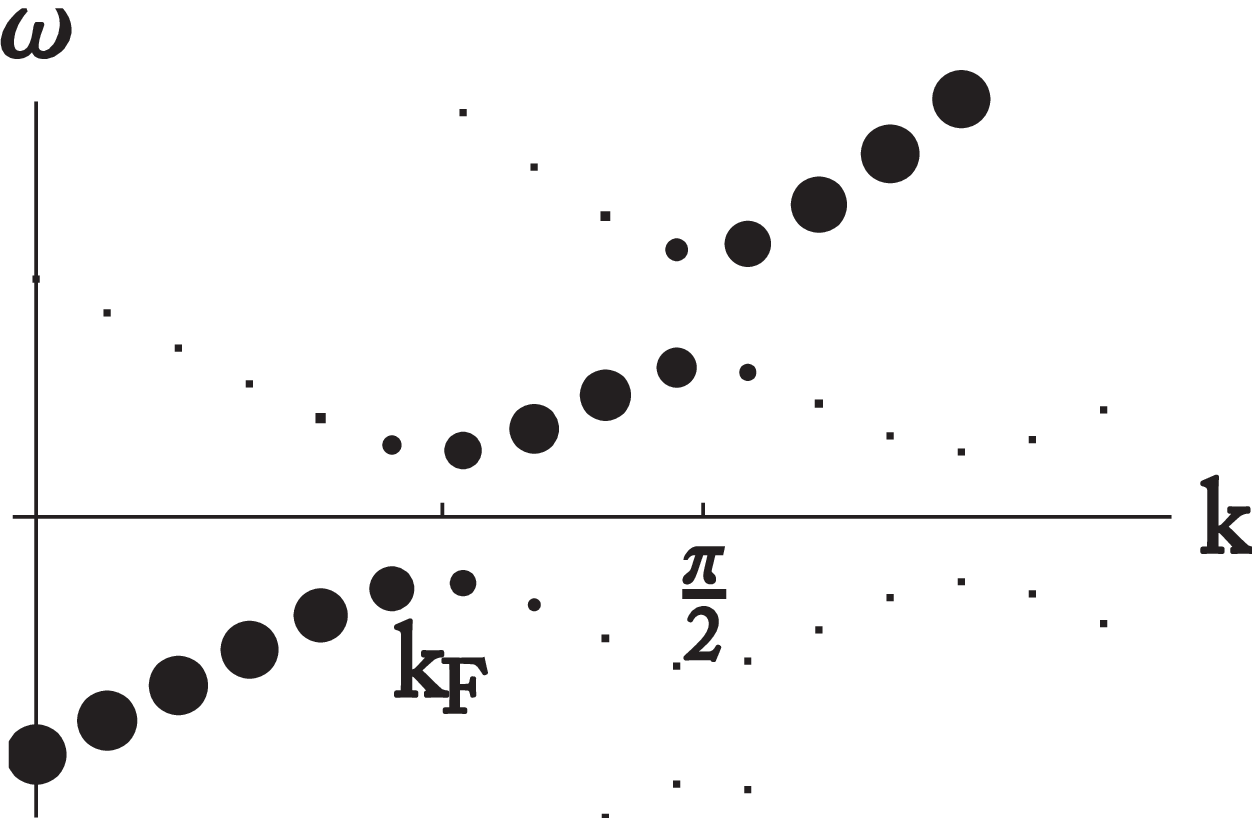} \end{minipage}
\begin{minipage}[br]{4.25cm} \includegraphics[width=1.4in]{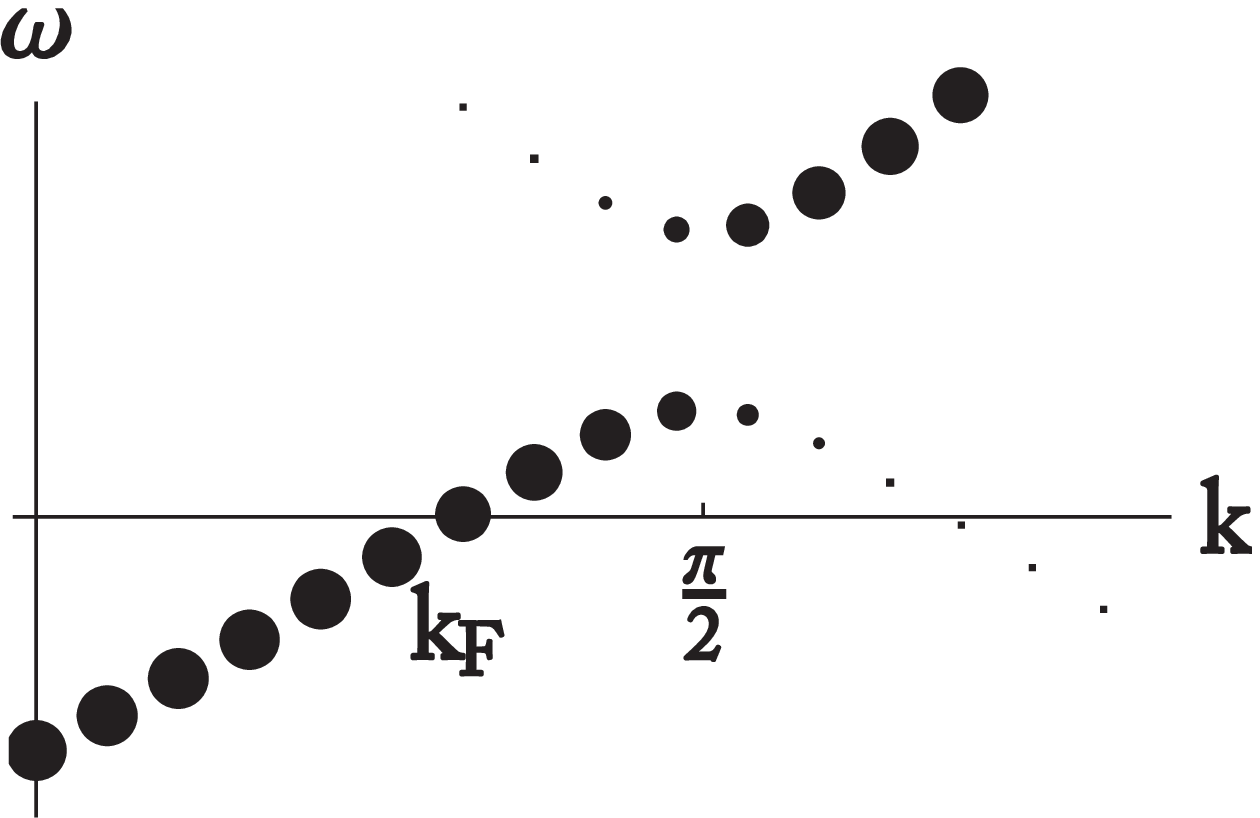} \end{minipage}
\caption{Schematic evolution of the spectra along a cut parallel to the nodal direction, inside the pocket (through region~I in Fig.~\ref{fig:FSscheme}, e.g.\ cut $b$ in Fig.~\ref{fig:Ak0_avg}). The dot size is proportional to the spectral weight. For upper left, upper right, lower left, lower right we have $\cos\theta = 0, 1/3, 2/3, 1$. Upper left is the superconducting state, lower right is the staggered-flux state.}
\label{fig:spectraWeightsIn}
\end{figure}

\begin{figure}[htbp]
\begin{minipage}[l]{4.25cm} \includegraphics[width=1.4in]{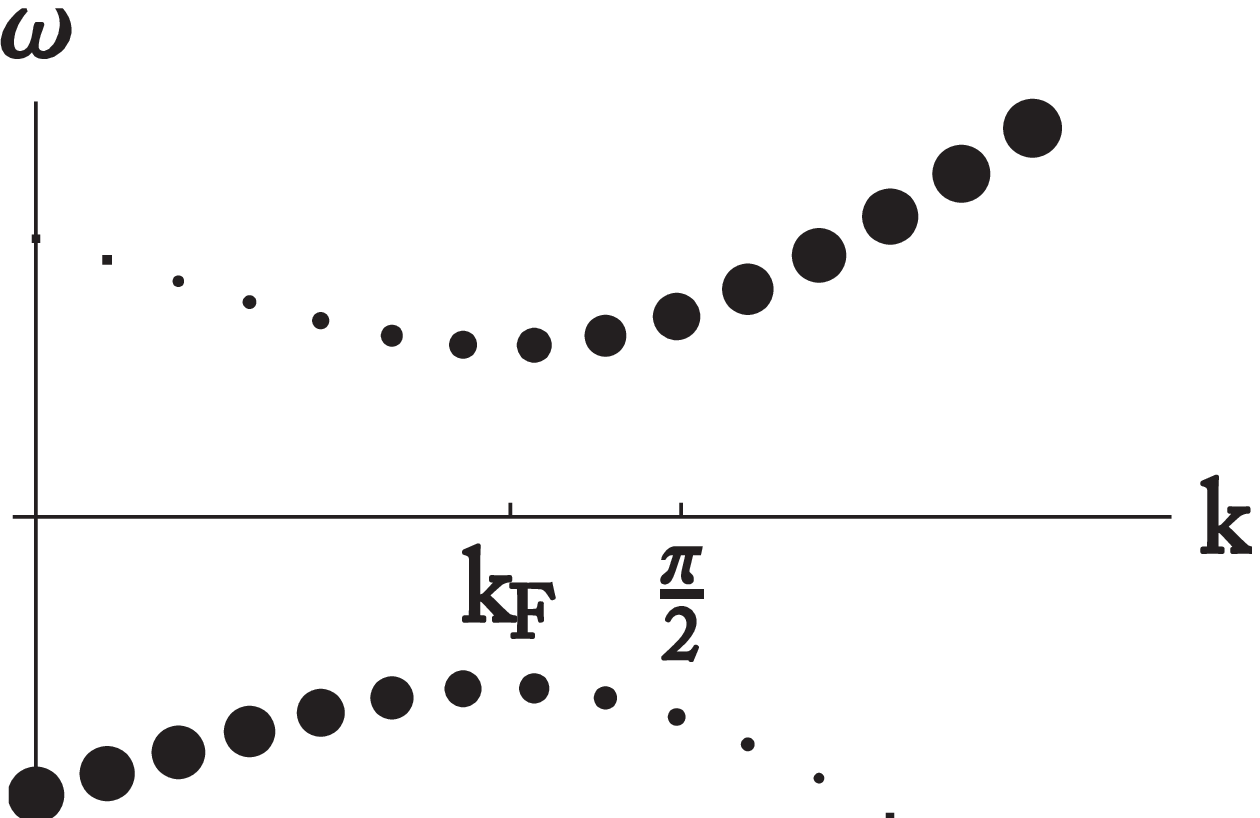} \end{minipage}
\begin{minipage}[c]{4.25cm} \includegraphics[width=1.4in]{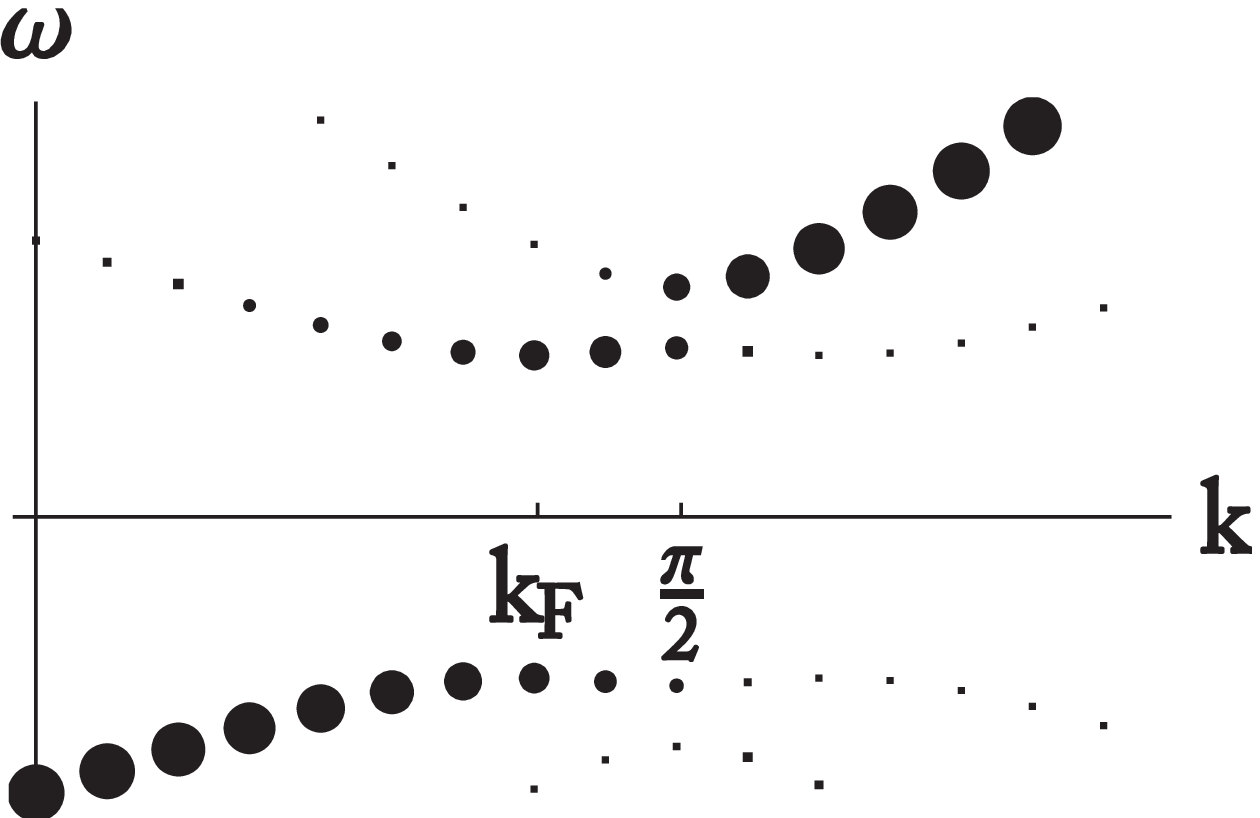} \end{minipage}
\begin{minipage}[r]{4.25cm} \includegraphics[width=1.4in]{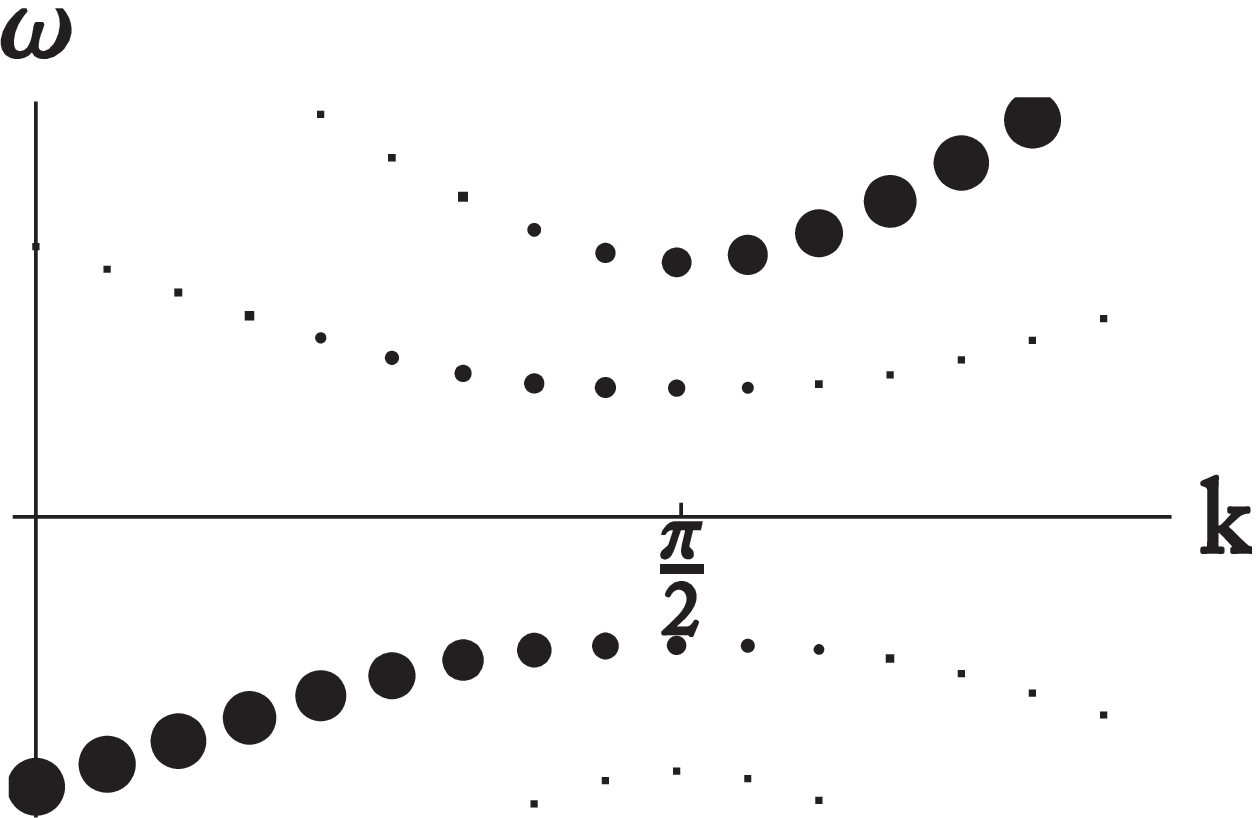} \end{minipage}
\begin{minipage}[r]{4.25cm} \includegraphics[width=1.4in]{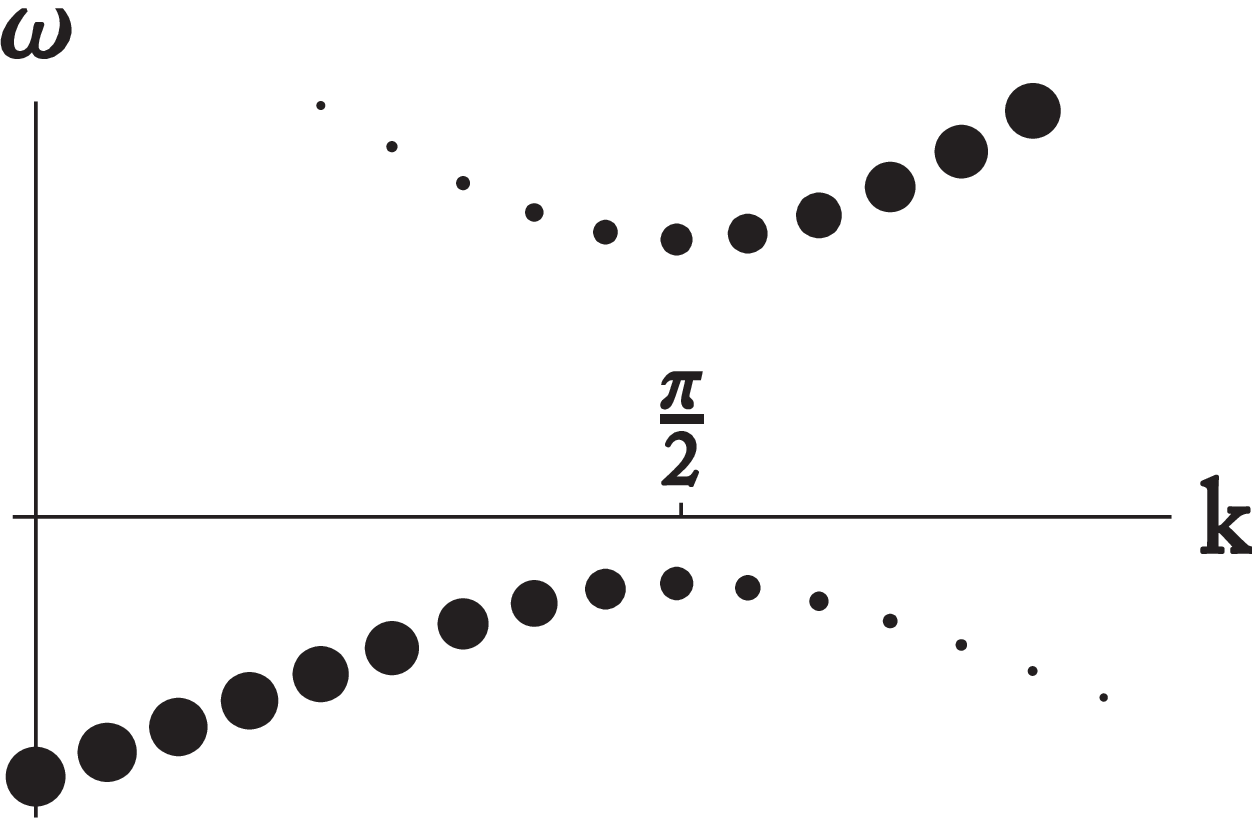} \end{minipage}
\caption{Same plot as in Fig.~\ref{fig:spectraWeightsIn}, but for a cut outside the pocket (through region~II in Fig.~\ref{fig:FSscheme}, e.g.\ cut $c$ in Fig.~\ref{fig:Ak0_avg}).}
\label{fig:spectraWeightsOut}
\end{figure}

\begin{figure}[htbp]
\begin{minipage}[l]{4.25cm} \includegraphics[width=1.4in]{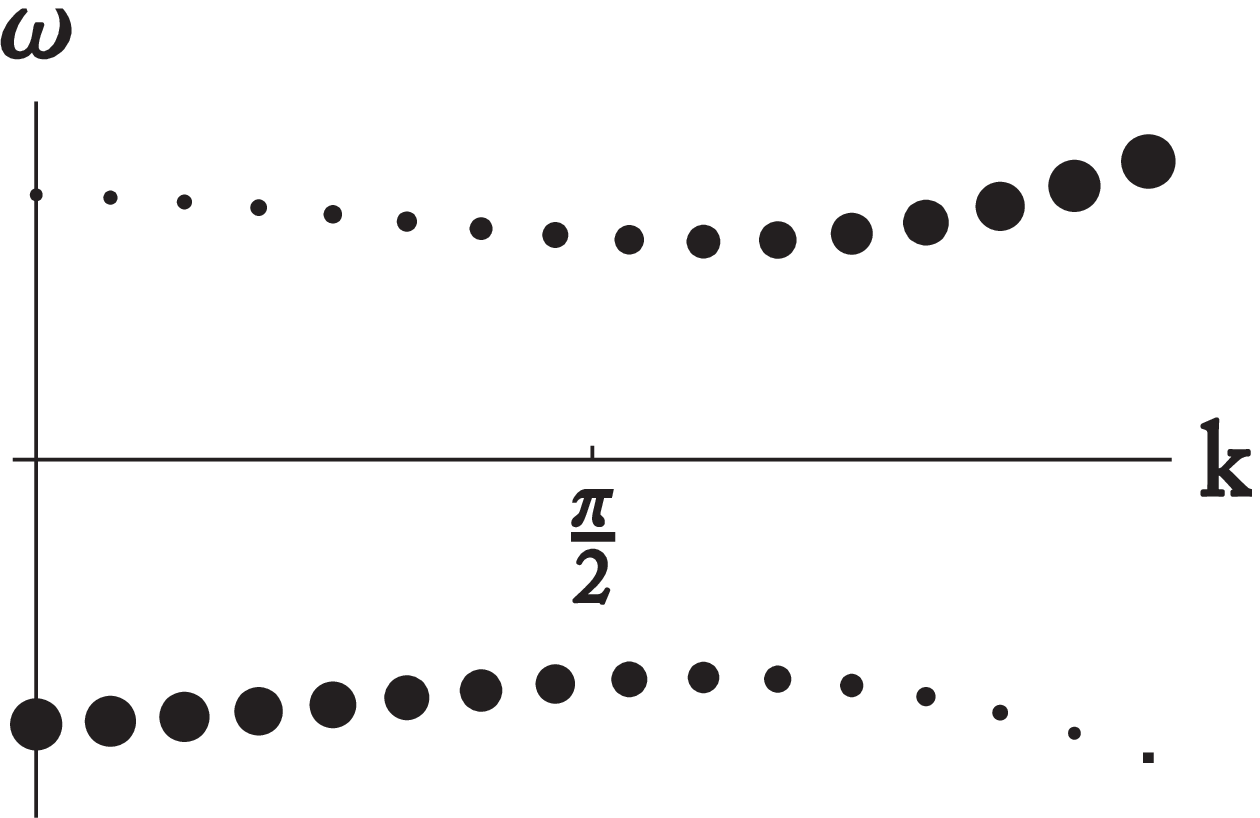} \end{minipage}
\begin{minipage}[c]{4.25cm} \includegraphics[width=1.4in]{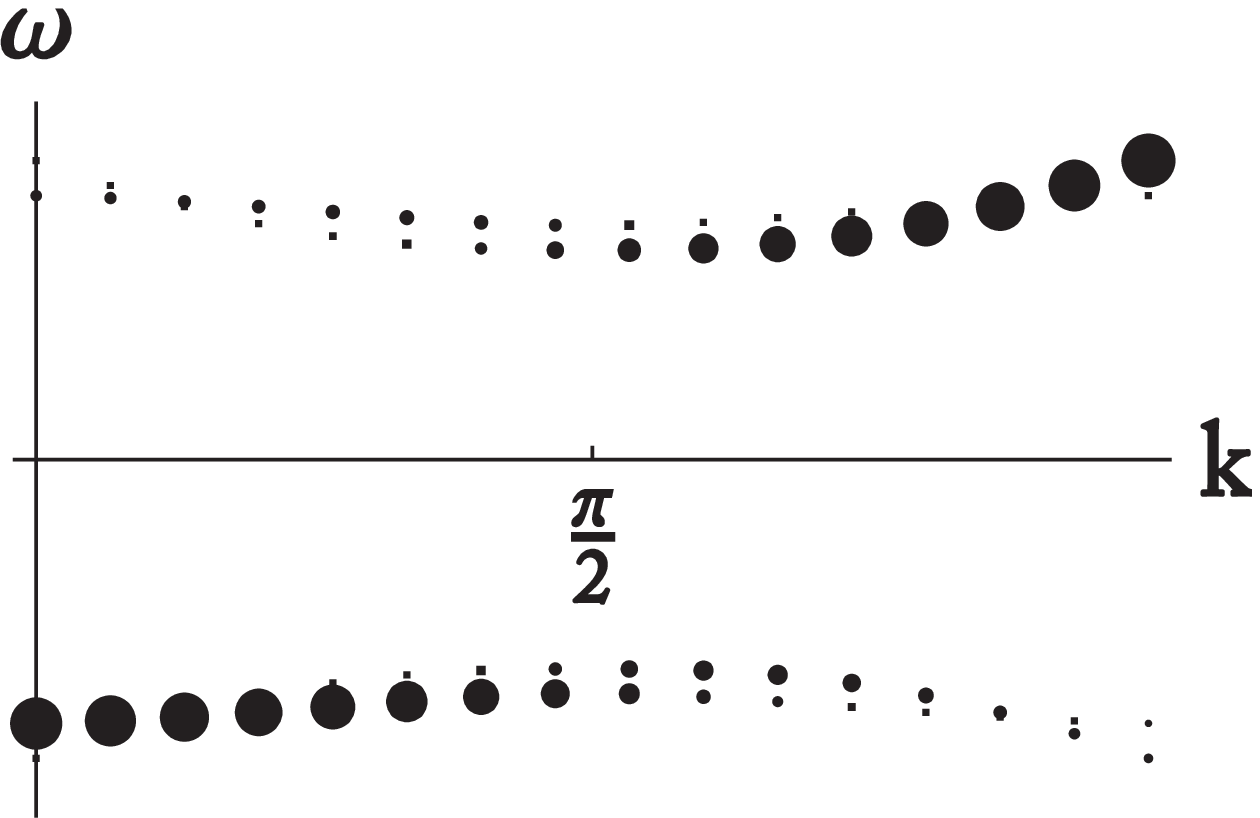} \end{minipage}
\begin{minipage}[r]{4.25cm} \includegraphics[width=1.4in]{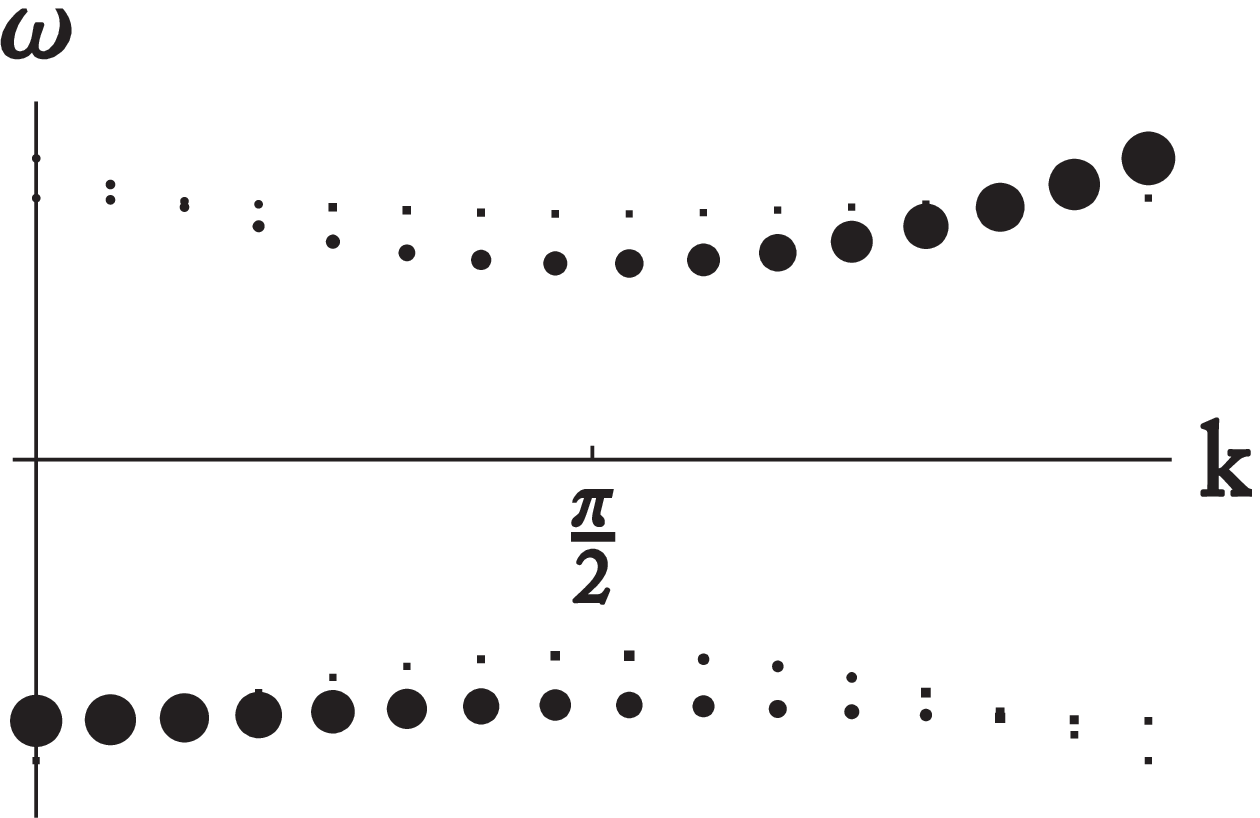} \end{minipage}
\begin{minipage}[r]{4.25cm} \includegraphics[width=1.4in]{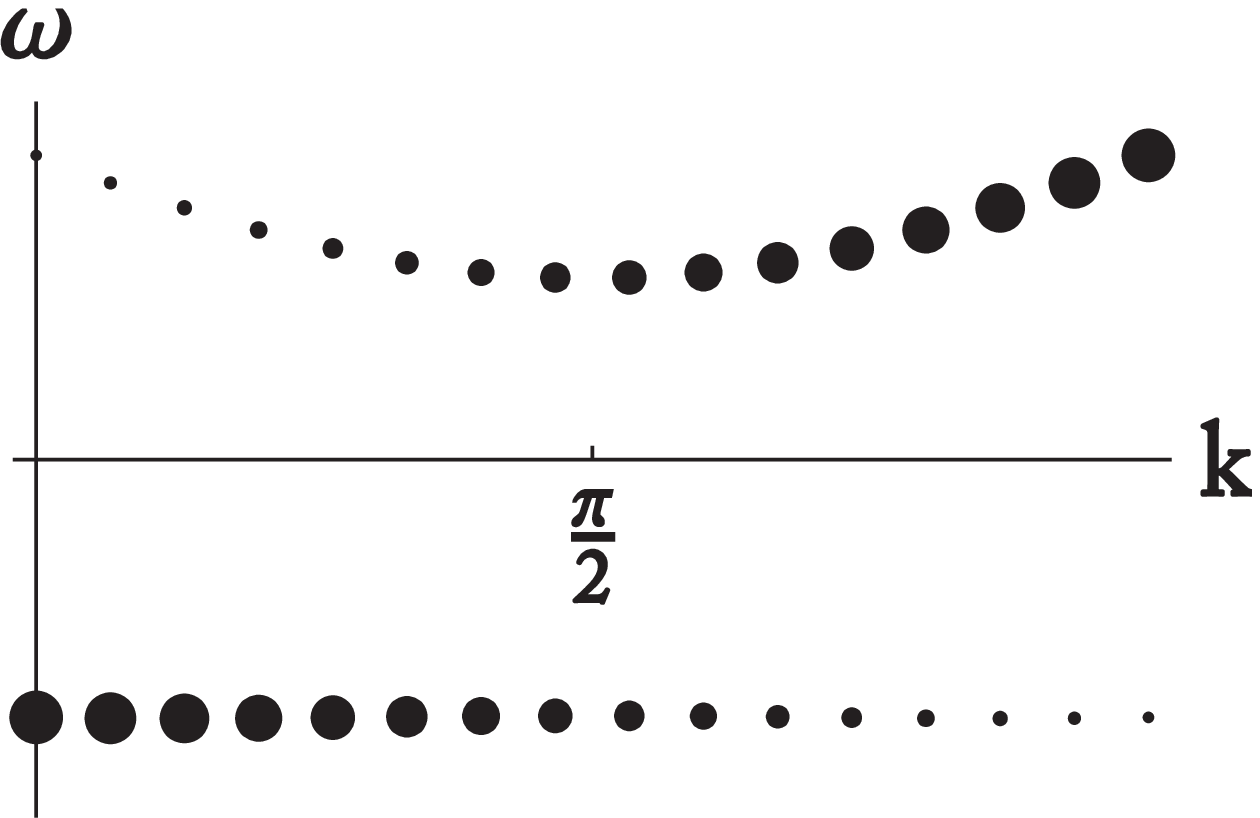} \end{minipage}
\caption{Same plot as in Fig.~\ref{fig:spectraWeightsIn}, but for a cut outside the SU(2)~point (through region~III in Fig.~\ref{fig:FSscheme}, e.g.\ cut $d$ in Fig.~\ref{fig:Ak0_avg}).}
\label{fig:spectraWeightsOut2}
\end{figure}

A schematic plot of the band structure and an illustration of the spectral-weight transfer as we go from the SC to the SF state is shown in Figs.~\ref{fig:spectraWeightsIn}, \ref{fig:spectraWeightsOut}, and \ref{fig:spectraWeightsOut2} on cuts parallel to the diagonal $(0,0)$-$(\pi,\pi)$. The behavior is qualitatively similar for all parallel cuts. The strong weights stay on the respective bands as they continuously move, except in a small stripe between the diagonal $(0,\pi)$-$(\pi,0)$ and the superconducting Fermi surface, outside the SF pocket (regions~II and III in Fig.~\ref{fig:FSscheme}). In region~II ($\tilde\mu<0$), the strong SC band at positive energy transfers some of its weight to the SF band at negative energy (see Fig.~\ref{fig:spectraWeightsOut}). Here, the midpoint of the SF bands lies at positive energy. In region~III ($\tilde\mu>0$), the strong SC band at negative energy transfers its weight to the SF band at positive energy (see Fig.~\ref{fig:spectraWeightsOut2}). The midpoint of the SF bands is now shifted below the Fermi energy.

\section{Averaged state}\label{sec:avg}

The gap considerations in Sec.~\ref{sec:pure} help now to understand the spectral properties of the averaged pseudogap state.
%
%
If we neglect the weak dependency $\Delta(\theta)$, the average gaps in the pocket region (region~I in Fig.~\ref{fig:FSscheme}) may be estimated as $\langle\Delta_{SC}\rangle\simeq\frac{\pi}{2}\Delta\,(\cos k_x-\cos k_y)$ and $\langle\Delta_{SF}\rangle\simeq\Delta\, (\cos k_x-\cos k_y)$. In the region outside the pocket,
the midgap energy of an effective gap is approximately given by $-\tilde\mu/2$.
%
%
On the other hand, if we are strict with the definition of the effective gap and only consider the truly excitation-free region, then we come to a picture with a gapless arc in region~I and an opening of an effective gap when the two gaps start to overlap in region~II. This effective gap opens above the Fermi energy and comes down as we move toward $(\pi,0)$. At the SU(2)~point, it is symmetric around the Fermi energy. Moving further out in region~III, we find an effective gap with midgap below the Fermi energy (see Fig.~\ref{fig:FSscheme}).

\begin{figure}[htbp]
\includegraphics[width=.45\textwidth]{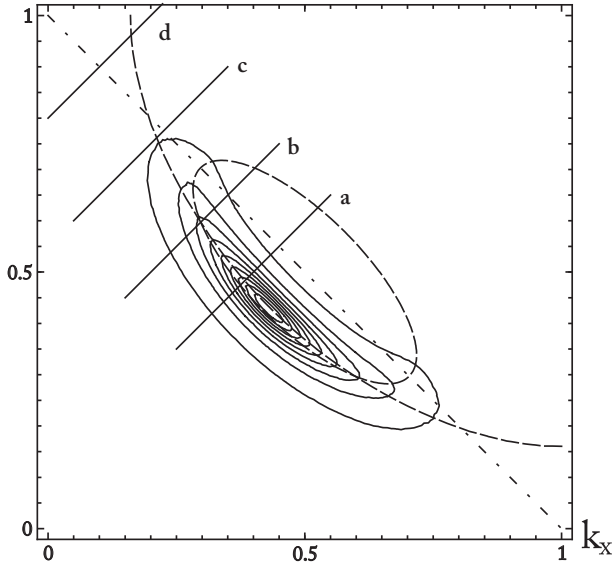}
\caption{Averaged spectral function at the Fermi energy, $A_{\bm k}(0)$. Doping is $10\%$ and we use a lifetime broadening $\Gamma=0.2\chi$. The dashed lines are the Fermi surfaces of the SF and SC states, respectively. The full spectra on the cuts $a$ to $d$ are given in Figs.~\ref{fig:Ak_cut1_avg}-\ref{fig:Ak_cut4_avg}.}
\label{fig:Ak0_avg}
\end{figure}

In Fig.~\ref{fig:Ak0_avg} we plot the averaged spectral function at the Fermi energy, $A_{\bm k}(0)$, with a quasiparticle lifetime broadening of $\Gamma=0.2\chi$. This spectral intensity resembles the superconducting one, in contrast to the pure SF pocket [see Fig.~\ref{fig:pureLS}(d)]. In fact, the absence of pocket-like features (``turn in" of the arc) in ARPES measurements was used in Ref.~\onlinecite{norman07} as an argument against the (static) staggered-flux state. However, in our averaged state this ``turn in" is completely washed out by the fluctuations toward the SC state, and the Fermi arc closely follows the SC Fermi surface, consistent with ARPES data.
%
%
%

\begin{figure}[htbp]
\includegraphics[width=.48\textwidth]{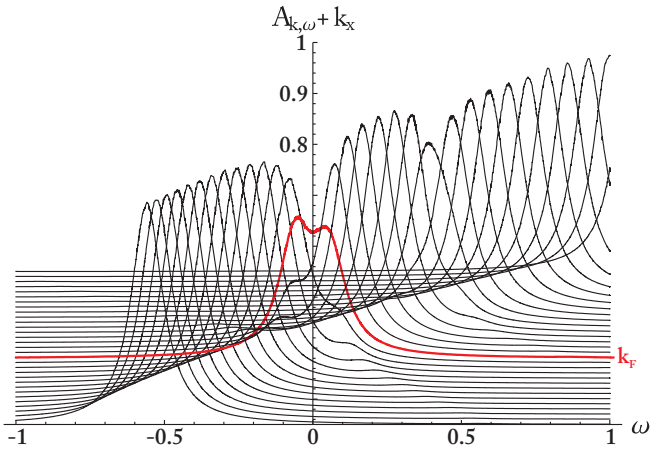}
\caption{(Color online) Averaged spectral function along a cut parallel to the nodal direction in the Brillouin zone, cut $a$ in Fig.~\ref{fig:Ak0_avg}. The spectra are set off in $y$ direction by $k_x$ (in units of $\pi$). $k_x$ goes from $0.25$ (lowest curve) to $0.55$ (highest curve) in steps of $0.01$. The curve at approximately $k_F$ is indicated on the right (red online). We use a lifetime broadening $\Gamma=0.12\chi$. The parameters used are $\Delta(\theta)=\sqrt{(0.2 \cos\theta)^2 + (0.25 \sin\theta)^2}$ and doping is $10 \%$. The energy~$\omega$ is given in units of $2 \chi\simeq 200\, {\rm meV}$, the intensities are in arbitrary units.
%
}
\label{fig:Ak_cut1_avg}
\end{figure}

\begin{figure}[htbp]
\includegraphics[width=.48\textwidth]{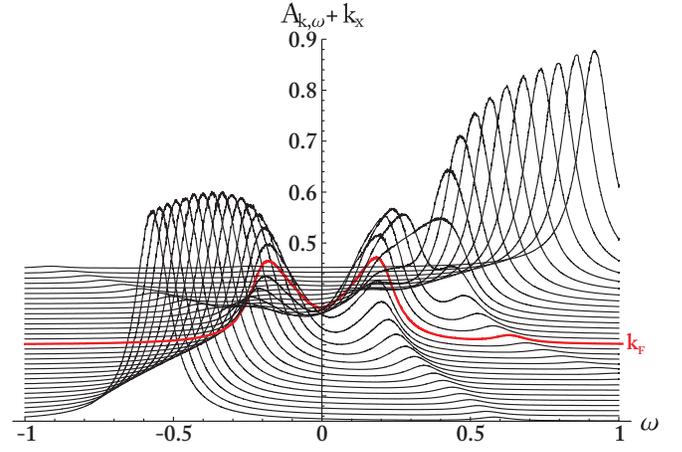}
\caption{(Color online) Same plot as in Fig.~\ref{fig:Ak_cut1_avg} but on cut $b$ in Fig.~\ref{fig:Ak0_avg}. $k_x$ goes from $0.15$ to $0.45$.}
\label{fig:Ak_cut2_avg}
\end{figure}

\begin{figure}[htbp]
\includegraphics[width=.48\textwidth]{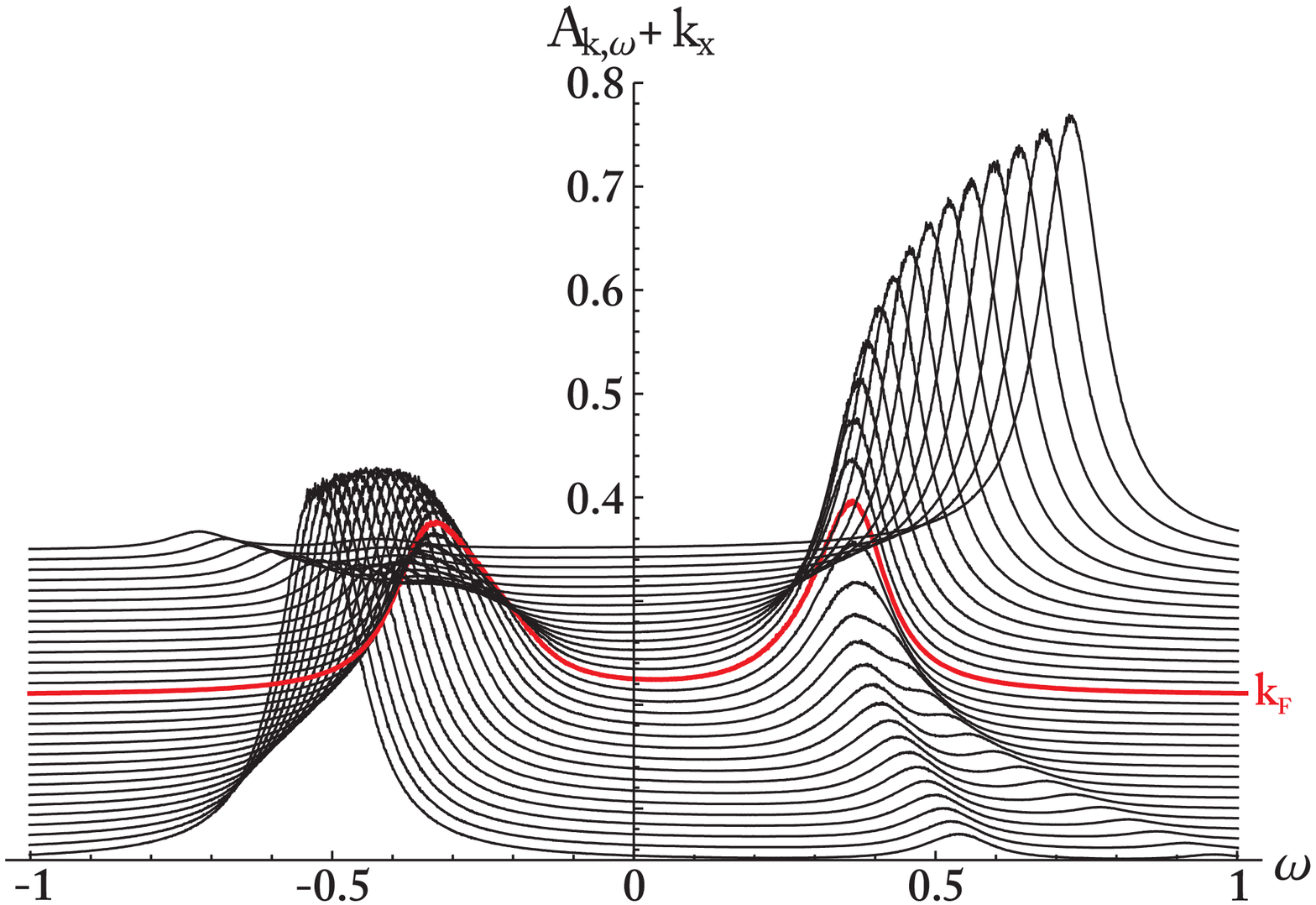}
\caption{(Color online) Same plot as in Fig.~\ref{fig:Ak_cut1_avg} but on cut $c$ in Fig.~\ref{fig:Ak0_avg}. $k_x$ goes from $0.05$ to $0.35$. An asymmetric effective gap with midgap above the Fermi energy is formed.}
\label{fig:Ak_cut3_avg}
\end{figure}

\begin{figure}[htbp]
\includegraphics[width=.48\textwidth]{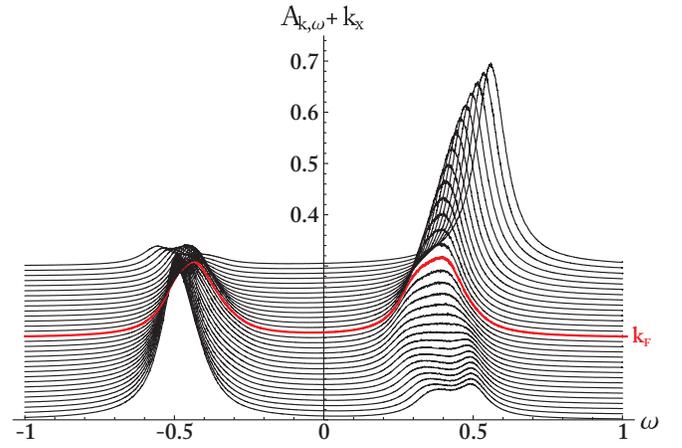}
\caption{(Color online) Same plot as in Fig.~\ref{fig:Ak_cut1_avg} but on cut $d$ in Fig.~\ref{fig:Ak0_avg}. $k_x$ goes from $0$ to $0.3$. An asymmetric effective gap with midgap below the Fermi energy is formed.}
\label{fig:Ak_cut4_avg}
\end{figure}


\clearpage
The averaged spectra on the cuts $a$ to $d$ in Fig.~\ref{fig:Ak0_avg} are shown in Figs.~\ref{fig:Ak_cut1_avg}-\ref{fig:Ak_cut4_avg}. In addition to the intrinsic width of the averaged state, we have chosen a lifetime broadening of $\Gamma=0.12\chi$ in these plots. From the averaged spectral function we can confirm what was already anticipated from the pure states:
\begin{enumerate}[(i)]
\item  In the region near the node, one can see a small symmetric suppression of intensity coming from the superconducting gap centered at the Fermi energy and a very small suppression coming from the staggered-flux gap centered above the Fermi energy. These ``gaps" are easily washed out by broadening effects (see Figs.~\ref{fig:Ak_cut1_avg} and \ref{fig:Ak_cut2_avg}) and may give rise to a Fermi arc.
\item Outside the arc, the (pseudo-)gap opens asymmetrically, with midgap first above the Fermi energy (Fig.~\ref{fig:Ak_cut3_avg}). Closer to the Brillouin-zone boundary, as we cross the SU(2)~point, the gap becomes asymmetric with midgap below the Fermi energy (Fig.~\ref{fig:Ak_cut4_avg}). At the SU(2)~point, the gap is exactly symmetric (see illustration in Fig.~\ref{fig:FSscheme}).
\item The backbending spectra at the edges of the two gaps lead to a doubling of the bands in some locations of the Brillouin zone (see Figs.~\ref{fig:Ak_cut2_avg} and \ref{fig:Ak_cut3_avg}). This band doubling only happens for weak branches and at positive energy.
\end{enumerate}

Finally, let us emphasize that the asymmetry we find here is in the location of the two pseudogap coherence peaks with respect to the Fermi energy. A different asymmetry in the renormalization of the coherent spectral weights in the superconducting state at low doping has been reported in recent variational Monte Carlo calculations, where the Gutzwiller constraint $n_i<2$ is taken into account exactly.\cite{bieri07,tanWang08,yangLi06} We expect that such a spectral-weight asymmetry is also present in our model (if one includes the Gutzwiller projection), but a confirmation would require an extensive numerical work.


\section{Experimental implications}\label{sec:exp}



The most striking prediction of our model, the formation of a staggered-flux gap above the Fermi energy, is difficult to verify directly in ARPES experiments, because this effect only appears at positive energy, around $\omega\simeq 100\, {\rm meV}$. On the other hand, our more subtle prediction, the combination of superconducting and staggered-flux gaps into a single asymmetric gap, appearing in the antinodal region of the cuprates may well be within current experimental reach. However, it is clear that the widely applied energy-symmetrization of the photoemission intensities~\cite{campuzano04,damascelli03} inevitably destroys all such signs in the ARPES spectra. A careful explicit removal of temperature- and device-dependent factors from the ARPES intensity will be extremely important in order to detect these effects. We hope that our work will stimulate experimental and theoretical effort in this direction.


{\it Note added.} After submission of this work, a possibly relevant ARPES study of the pseudogap phase of underdoped Bi2212 was published.\cite{yangJohnson08} These authors found a gap asymmetry in the experimental spectral function which is in agreement with our findings. An alternative approach to ours was able to fit these data with a phenomenological spectral function.\cite{yangRiceZhang08}

\begin{acknowledgments}
We would like to thank Patrick A.\ Lee, Tao Li, Alexander G.\ Abanov, George Jackeli, Mike R.\ Norman, and Davor Pavuna for helpful discussions. This work was supported by the Swiss National Science Foundation.
\end{acknowledgments}



\begin{thebibliography}{99}
\bibitem{note_params}
For the averaging, we interpolate the parameter $\Delta$ in $\theta$ as $\Delta(\theta)=\sqrt{\Delta_{\theta=0}^2 (\cos\theta)^2 + \Delta_{\theta=\pi/2}^2 (\sin\theta)^2}$.
The weak dependence of $\Delta$ on $\theta$ is not important for the main conclusions of our study.
Note that we are using the mean-field order parameters at zero temperature for our calculation. In order to reliably estimate the temperature and doping dependence of the spectral function, our work would need a considerable extension, e.g.\ by a self-consistent determination of the mean-field order parameters at finite temperature. We leave this for future research.
\bibitem{note_rho}
For simplicity, we do not distinguish between the two stiffnesses: for $\theta$ and for the superconducting phase. We assume them to be of the same order of magnitude and much larger than $\varepsilon_c$. This assumption does not fully agree with earlier numerical calculations with Gutzwiller-projected wavefunctions (where the superfluid stiffness was estimated to have the same order of magnitude as $\varepsilon_c$),\cite{ivanovLee03} but is consistent with the experimental estimates of $\xi$ of order 5-10 lattice spacings.\cite{cohlength}
\bibitem{note_gf}
Note that the unit cell for $H_{MF}$ [Eq.~\eqref{eq:Hmf}] contains two sites. Therefore, one may define the normal Green's function as a 2$\times$2 matrix. However, for comparison with the ARPES intensity, we are considering here the Green's function $G_{\bm k}(t) = i \theta(t) \langle \{ c_{\bm k}(t), c^\dagger_{\bm k}(0) \} \rangle$ which corresponds to the sum over all entries of the matrix Green's function.\\
Since $H_{MF}$ is a quadratic model, $G_{\bm k}(\omega)$ [and the spectral function $A_{\bm k}(\omega) = -\frac{1}{\pi}\text{Im} G_{\bm k}(\omega)$] is independent of temperature. The intensity measured in direct photoemission at finite temperature is given by the spectral function times the Fermi distribution: $I({\bm k},\omega) \propto [1+\exp{\beta\omega}]^{-1} A_{\bm k}(\omega)$.\cite{campuzano04,damascelli03}
\bibitem{note_FS}
Known subtleties and discrepancies in the various definitions of the experimental Fermi surface\cite{grosAnderson06,sensarmaRanderia07} are not in the scope of this paper; here, we use the theoretically well-defined Luttinger surface where the mean-field Green's function changes sign, $G_{{\bm k}_F}(0)=0,\pm\infty$.\cite{dzyaloshinksii03} The Green's function described in Remark~\onlinecite{note_gf} is used.
\end{thebibliography}
\end{document}